\documentclass[11pt]{article}%%%%where rstrans is the template name
\usepackage{geometry}
\usepackage{cite}
\usepackage{graphicx}
\usepackage{amssymb}
\usepackage{epstopdf}
\newcommand{\be}{\begin{eqnarray}}
\newcommand{\ee}{\end{eqnarray}}

\newcommand{\bc}{\begin{center}}
\newcommand{\ec}{\end{center}}
\begin{document}

%%%% Article title to be placed here
\title{What is Information?\footnote{The present comment was adapted from a blog series by the author, with the same title. It is meant to be an elementary introduction to information theory for non-experts that assumes no prior exposure to the subject. The tone is deliberately conversational, and no effort is made to cite the primary literature. }}

\author{%%%% Author details
Christoph Adami
}
\date{}
\maketitle
%%%%%%%%% Insert author address here
\begin{center}{Department of Microbiology and Molecular Genetics\\
Department of Physics and Astronomy\\
BEACON Center for the Study of Evolution in Action\\
Michigan State University, East Lansing, MI 48824\\
email: adami@msu.edu}
\end{center}
%%%% Subject entries to be placed here %%%%
%\subject{information theory}

%%%% Keyword entries to be placed here %%%%
%\keywords{entropy, information}

%%%% Insert corresponding author and its email address}
%\corres{C.Adami~\email{adami@msu.edu}}

%%%% Abstract text to be placed here %%%%%%%%%%%%
\begin{abstract}
Information is a precise concept that can be defined mathematically, but its relationship to what we call ``knowledge" is not always made clear. Furthermore, the concepts ``entropy" and ``information", while deeply related, are distinct and must be used with care, something that is not always achieved in the literature. In this elementary introduction, the concepts of entropy and information are laid out one by one, explained intuitively, but defined rigorously. I argue that a proper understanding of information in terms of prediction is key to a number of disciplines beyond engineering, such as physics and biology.
\end{abstract}
%%%%%%%%%%%%%%%%%%%%%%%%%%%

%%%%%%%%%% Insert the texts which can accomdate on firstpage in the tag "fmtext" %%%%%

\section{Entropy: In the Eye of the Beholder}
Information is a central concept in our daily life. We rely on information in order to make sense of the world: to make ``informed" decisions. We use information technology in our daily interactions with people and machines. Even though most people are perfectly comfortable with their day-to-day understanding of information, the precise definition of information, along with its properties and consequences, is not always as well understood. I want to argue in this comment that a precise understanding of the concept of information is crucial to a number of scientific disciplines. Conversely, a vague understanding of the concept can lead to profound misunderstandings, within daily life and within the technical scientific literature.  My purpose is to introduce the concept of information--mathematically defined--to a broader audience, with the express intent of eliminating a number of common misconceptions that have plagued the progress of information science in different fields.

What is information? Simply put, information is that which allows you (who is in possession of that information) to make predictions with accuracy better than chance. Even though the former sentence appears glib, it captures the concept of information fairly succinctly. But the concepts introduced in this sentence need to be clarified. What do I mean with prediction? What is ``accuracy better than chance"? Predictions of what? 

We all understand that information is useful. When is the last time that you have found information to be counterproductive? Perhaps it was the last time you watched the News. I will argue that, when you thought that the information you were given was not useful, then what you were exposed to was most likely not information. That stuff, instead, was mostly entropy (with a little bit of information thrown in here or there). Entropy, in case you have not yet come across the term,  is just a word we use to quantify how much isn't known.

\vskip 0.5cm
\bc
{\em ``But, isn't entropy the same as information?"}
\ec
%\vskip -0.3cm
One of the objectives of this comment is to make the distinction between the two as clear as possible. Information and entropy are two very different objects. They may have been used synonymously (even by Claude Shannon--the father of information theory--thus being responsible in part for a persistent myth) but they are fundamentally different. If the only thing you will take away from this article is your appreciation of the difference between entropy and information, then I will have succeeded.

But let us go back to our colloquial description of what information is, in terms of predictions. ``Predictions of what"? you should ask. Well, in general, when we make predictions, they are about a system that we don't already know. In other words, an {\em other} system. This other system can be anything: the stock market, a book, the behavior of another person. But I've told you that we will make the concept of information mathematically precise. In that case, I have to specify this ``other system" as precisely as I can. I have to specify, in particular, which states the system can take on. This is, in most cases, not particularly difficult. If I'm interested in quantifying how much I don't know about a phone book, say, I just need to tell you the number of phone numbers in it. Or, let's take a more familiar example (as phone books may appeal, conceptually, only to the older crowd among us), such as the six-sided fair die. What I don't know about this system is which side is going to be up when I throw it next. What you do know is that it has six sides. How much don't you know about this die? The answer is not six. This is because information (or the lack thereof) is not defined in terms of the number of unknown states. Rather, it is given by the {\em logarithm} of the number of unknown states. 
\vskip 0.5cm
\bc
{\em  ``Why on Earth introduce that complication?"}, you ask.
\ec
%\vskip 0.5cm
Well, think of it this way. Let's quantify your uncertainty (that is, how much you don't know) about a system (System One) by the number of states it can be in. Say this is $N_1$. Imagine that there is another system (System Two), and that one can be in $N_2$ different states. How many states can the joint system (System One And Two Combined) be in? Well, for each state of System One, there can be $N_2$ number of states. So the total number of states of the joint system must be $N_1\times N_2$. But our uncertainty about the joint system is not $N_1\times N_2$. Our uncertainty adds, it does not multiply. And fortunately the logarithm is that one function where the log of a product of elements is the sum of the logs of the elements. So, the uncertainty (let's call it $H$) about the system $N_1\times N_2$ is the logarithm of the number of states
$$H(N_1N_2)=\log(N_1N_2)=\log(N_1) + \log(N_2).$$
%I had to assume here that you knew about the properties of the log function. If this is a problem for you, please consult Wikipedia and continue after you digested that content.
%Phew, I'm glad we got this out of the way. 
Let's return to the six-sided die. You know, the type you've known most of your life. What you don't know about the state of this die (your uncertainty) {\em before} throwing the die is $\log 6$. When you peek at the number that came up, you have reduced your uncertainty (about the outcome of this throw) to zero. This is because you made a perfect measurement. (In an imperfect measurement, you only got a glimpse of the surface that rules out a ``1" and a ``2", say.) 

What if the die wasn't fair? Well that complicates things. Let us for the sake of the argument assume that the die is so unfair that one of the six sides (say, the ``six") can never be up. You might argue that the {\em a priori} uncertainty of the die (the uncertainty before measurement) should now be $\log 5$, because only five of the states can be the outcome of the measurement. But how are you supposed to know this? You were not told that the die is unfair in this manner, so as far as you are concerned, your uncertainty is still $\log 6$. 

Absurd, you say? You say that the entropy of the die is whatever it is, and does not depend on the state of the observer? Well I'm here to say that if you think that, then you are mistaken. Physical objects do not have an intrinsic uncertainty. I can easily convince you of that. You say the fair die has an entropy of $\log 6$? Let's look at an even more simple object: the fair coin. Its entropy is $\log 2$, right? What if I told you that I'm playing a somewhat different game, one where I'm not just counting whether the coin comes up heads or tails, but am also counting the angle that the face has made with a line that points towards True North. And in my game, I allow four different quadrants, like in Fig.~\ref{fig:coin} below.

\begin{figure}[htbp] %  figure placement: here, top, bottom, or page
   \centering
   \includegraphics[width=4in]{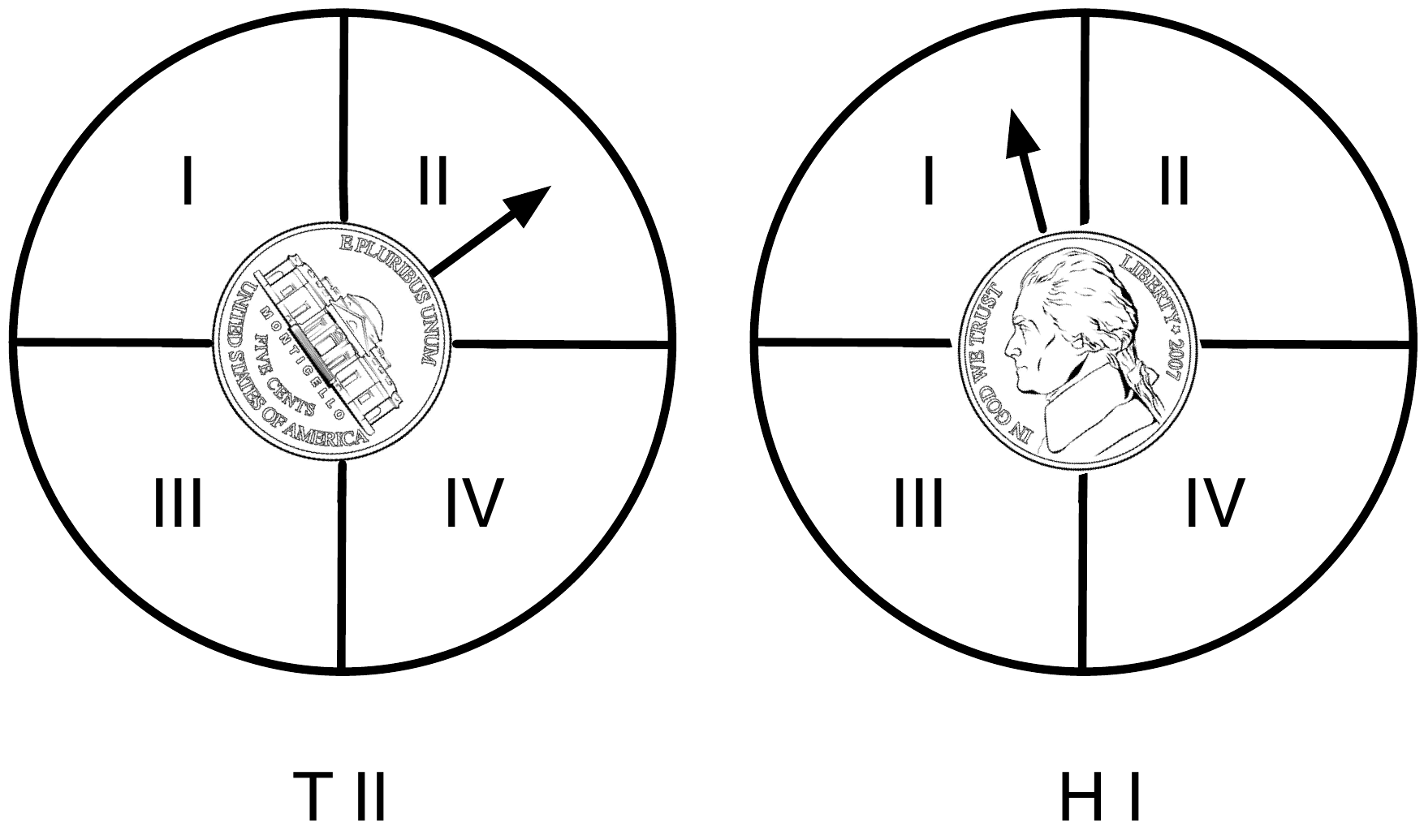} 
   \caption{A fair coin with entropy $\log_2 8=3$ bits. In the figure on the left, the outcome is ``Tails, quadrant II", while the coin on the right landed as ``Heads, quadrant I".  }
   \label{fig:coin}
\end{figure}

Suddenly, the coin has $2\times4$ possible states, just because I told you that in my game the angle that the face makes with respect to a circle divided into 4 quadrants is interesting to me. It's the same coin, but I decided to measure something that is actually measurable (because the coin's faces can be in different orientation, as opposed to, say, a coin with a plain face but two differently colored sides). And you immediately realize that I could have divided the circle into as many quadrants as I can possibly resolve by eye. 

All right, fine, you say, so the entropy is $\log(2\times N)$ where $N$ is the number of resolvable angles. But you know, what is resolvable really depends on the measurement device you are going to use. If you use a microscope instead of your eyes, you could probably resolve many more states. Actually, let's follow this train of thought. Let's imagine I have a very sensitive thermometer that can sense the temperature of the coin. When throwing it high (the coin, not the thermometer) the energy the coin absorbs when hitting the surface will raise the temperature of the coin slightly, compared to a coin that was tossed gently. If I so choose, I could include this temperature as another characteristic, and now the entropy is $\log(2\times N\times M)$, where $M$ is the number of different temperatures that can be reliably measured by the device. And now you realize that I can drive this to the absurd, by deciding to consider the excitation states of the molecules that compose the coin, or of the atoms composing the molecules, or nuclei, the nucleons, or even the quarks and gluons. 

The entropy of a physical object, it dawns on you, is not defined unless {\em you} tell me which degrees of freedom are important to {\em you}. In other words, it is defined by the number of states that can be resolved by the measurement that you are going to be using to determine the state of the physical object. If it is heads or tails that counts for you, then $\log 2$ is your uncertainty. If you play the ``4-quadrant" game, the entropy of the coin is $\log 8$, and so on. Which brings us back to six-sided die that has been mysteriously manipulated to never land on ``six". You (who do not know about this mischievous machination) expect six possible states, so this dictates your uncertainty. Incidentally, how do you even know the die has six sides it can land on? You know this from experience with dice, and having looked at the die you are about to throw. This knowledge allowed you to quantify your a priori uncertainty in the first place. (I'll discuss prior knowledge in much more detail in the next section)

Now, you start throwing this weighted die, and after about twenty throws or so without a ``six" turning up, you start to become suspicious. You write down the results of a longer set of trials, and still note this curious pattern of ``six" {\em never} showing up, but you find that the other five outcomes occur with roughly equal frequency. What happens now is that you adjust your expectation. You now hypothesize (a posteriori) that it is a weighted die with five equally likely outcome, and one outcome that never actually occurs. Now your {\em expected} uncertainty is $\log 5$. (Of course, you can't be 100\% sure because you only took a finite number of trials.)

But you did learn something through all these measurements. You gained information. How much? Easy! It's the difference between your uncertainty before you started to be suspicious, minus the uncertainty after it dawned on you. The information you gained is just $\log 6-\log5$. How much is that? Well, you can calculate it yourself. You didn't give me the base of the logarithm you say? 
Well, that's true. Without specifying the logarithm's base, the information gained is not specified. It does not matter which base you choose: each base just gives units to your information gain. It's kind of like asking how much you weigh. Well, my weight is one thing. The number I give you depends on whether you want to know it in kilograms, or pounds. Or stones, for all it matters.

If you choose the base of the logarithm to be 2, well then your units will be called ``bits" (which is what we all use in information theory land). But you may choose the Eulerian $e$ as your base. That makes your logarithms "natural", but your units of information (or entropy, for that matter) will be called "nats".  You can define other units, but we'll keep it at that for the moment. 
So, if you choose base 2 (bits), your information gain is $\log_2(6/5)\approx 0.263$ bits. That may not sound like much, but in a Vegas-type setting this gain of information might be worth, well, a lot. Information that you have (and others do not) can be moderately valuable (for example, in a stock market setting), or it could mean the difference between life and death (in a predator/prey setting). In any case, we should value information.  

As an aside, this little example where we used a series of experiments to ``inform" us that one of the six sides of the die will not, in all likelihood, ever show up, should have convinced you that we can never know the {\em actual} uncertainty that we have about any physical object, unless the statistics of the possible measurement outcomes of that physical object are for some reason known with infinite precision (which you cannot attain in a finite lifetime). It is for that reason that I suggest to the reader to give up thinking about the uncertainty of any physical object, and be only concerned with differences between uncertainties (before and after a measurement, for example). The uncertainties themselves we call entropy. Differences between entropies (for example before and after a measurement) are called information. Information, you see, is real. Entropy on the other hand, is in the eye of the beholder.

%Part 2
\section{The Things We Know}
In the first section I have written mostly about entropy. How the entropy of a physical system (such as a die, a coin, or a book) depends on the measurement device that you will use for querying that system. That, come to think of it, the uncertainty (or entropy) of any physical object really is infinite, and made finite only by the finiteness of our measurement devices. 
Of course the things you could possibly know about any physical object is infinite! Think about it! Look at any object near to you. OK, the screen in front of you. Just imagine a microscope zooming in on the area framing the screen, revealing the intricate details of the material. The variations that the manufacturing process left behind, making each and every computer screen (or iPad or iPhone), essentially unique.
If this was a more formal article (as opposed to a comment), I would now launch into a discussion of how there is a precise parallel (really!) to renormalization in quantum field theory... but it isn't. So, let's instead delve head-first into the matter at hand, to prepare ourselves for a discussion of the concept of information.

What does it even mean to have information? Yes, of course, it means that you know something. About something. Let's make this more precise. I'll conjure up the old ``urn". The urn has things in it. You have to tell me what they are. So, now imagine that.....

\vskip 0.5cm
\bc
{\em ``Hold on, hold on. Who told you that the urn has things in it? Isn't that information already? Who told you that?"}
\ec
%\vskip 0.5cm

OK, fine, good point. But you know, the urn is really just a stand-in for what we call "random variables" in probability theory. A random variable is a ``thing" that can take on different states. Kind of like the urn, that you draw something from? When I draw a blue ball, say, then the ``state of the urn" is blue. If I draw a red ball, then the ``state of the urn" is red. So, ``urn=random variable". OK?
\vskip 0.5cm
\bc
{\em ``OK, fine, but you haven't answered my question. Who told you that there are blue and red balls in it? Who?"}
\ec
Let me think about that. Here's the thing. When a mathematician defines a random variable, they tell you which state it can take on, and with what probability. Like: ``A fair coin is a random variable with two states. Each state can be taken on with equal probability one-half." When they give you an urn, they also tell you how likely it is to get a blue or a red ball from it. They just don't tell you what you will {\em actually} get when you pull one out. 
\vskip 0.5cm
\bc
{\em ``But is this how real systems are? That you know the alternatives before asking questions?"}
\ec
%\vskip 0.5cm
All right, all right. I'm trying to teach you information theory, the way it is taught in any school you would set your foot in. I concede, when I define a random variable, then I tell you how many states it can take on, and what the probability is that you will see each of these states, when you ``reach into the random variable". Let's say that this info is magically conferred upon you. Happy now?
\vskip 0.5cm
\bc
{\em ``Not really."}
\ec
%\vskip 0.5cm
OK, let's just imagine that you spend a long time with this urn, and after a while of messing with it, you realize that:

\noindent A) This urn has balls in it.\\
\noindent B) From what you can tell, they are blue and red.\\
\noindent C) Reds occur more frequently than blues, but you're still working on what the ratio is.\\
Is this enough?
\vskip 0.5cm
\bc
{\em ``At least now we're talking. Do you know that you assume a lot when you say "random variable?"}
\ec
All right, you're making this more difficult than I intended it to be. According to standard lore, it appears that you're allowed to assume that you know something about the things you know nothing about. Let's just call these things ``common sense". Like, that a coin has two sides. Or an urn that has red and blue balls in it.  They could be any pair of colors, you do realize. And the things you {\em don't know} about the random variable are the things that go beyond common sense. The things that, unless you had performed dedicated experiments to ascertain the state of the variables, you wouldn't already know. 

How much don't you know about it?  Easily answered using our good buddy Shannon's insight. How much you don't know is quantified by the ``entropy" of the urn. That's calculated from the fraction of blue balls known to be in the urn, and  the fraction of red balls in the urn. You know, these fractions that are common knowledge. So, let's say that fraction of blue is $p$. The fraction of red then is of course (you do the math) $1-p$. And the entropy of the urn is
\be
                                H(X)=-p\log p-(1-p)\log(1-p)\;.          \label{ent}
\ee
\vskip 0.5cm
\bc
{\em In the first section you wrote that the entropy is $\log N$, where $N$ is the number of states of the system. Are you changing definitions on me?}
\ec
%\vskip 0.5cm
I'm not, actually. I just used a special case of the entropy to get across the point that the uncertainty/entropy is additive. It was the special case where each possible state occurs with equal likelihood. In that case, the probability $p$ is equal to $1/N$, and the above formula (\ref{ent})  turns into the first one. But let's get back to our urn. I mean random variable. And let's try to answer the question: 
\vskip 0.5cm
\bc
{"How much is there to know (about it)?"}
\ec
%\vskip 0.5cm
Assuming that we know the common knowledge stuff that the urn only has read and blue balls in it, then what we don't know is the identity of the next ball that we will draw. This drawing of balls is our experiment. We would love to be able to predict the outcome of this experiment exactly, but in order to pull off this feat, we would have to have some information about the urn. I mean, the contents of the urn. 

If we know nothing else about this urn, then the uncertainty is equal to the log of the number of possible states, as I wrote before. Because there are only red and blue balls, that would be log 2. And if the base of the log is two, then the result is $\log_2 2=1$ bit.  So, if there are red and blue balls only in an urn, then I can predict the outcome of an experiment (pulling a ball from the urn) just as well as I can predict whether a fair coin lands on heads or tails. If I correctly predict the outcome (I will be able to do this about half the time, on average) I am correct purely by chance. Information is that which allows you to make a correct prediction with accuracy better than chance, which in this case means, more than half of the time. 
\vskip 0.5cm
\bc
{\em ``How can you do this for the case of the fair coin, or the urn with equal numbers of red and blue balls?"}
\ec
Well, you can't unless you cheat. I should say, the case of the urn and of the fair coin are somewhat different. For the fair coin, I could use the knowledge of the state of the coin before flipping, and the forces acting on it during the flip, to calculate how it is going to land, at least approximately. This is a sophisticated way to use extra information to make predictions (the information here is the initial condition of the coin) but something akin to that has been used by a bunch of physics grad students to predict the outcome of casino roulette in the late 70s (you can read about it in~\cite{Bass1985}.) 

The coin is different from the urn because for the urn you won't be able to get any ``extraneous" information. But suppose the urn has blue and red balls in {\em unequal} proportions. If you knew what these proportions were [the $p$ and $1-p$ in Eq. (\ref{ent}) above] then you could reduce the uncertainty of 1 bit to $H(X)$. A priori (that is, before performing any measurements on the probability distribution of blue and read balls), the distribution is of course given by $p=1/2$, which is what you have to assume in the absence of information. That means your uncertainty is 1 bit. But keep in mind (from section 1)  that it is only one bit because you have decided that the color of the ball (blue or red) is what you are interested in predicting.

If you start drawing balls from the urn (and then replacing them, and noting down the result, of course) you would be able to estimate $p$ from the frequencies of blue and red balls. So, for example, if you end up seeing 9 times as many red balls as blue balls, you should adjust your prediction strategy to ``The next one will be red". And you would likely be right about 90\% of the time, quite a bit better than the 50/50 prior.
\vskip 0.5cm
\bc
{\em ``So what you are telling me, is that the entropy formula (\ref{ent}) assumes a whole lot of things, such as that you already know to expect a bunch of things, namely what the possible alternatives of the measurement are, and even what the probability distribution is, which you can really only know if you have divine inspiration, or else made a ton of measurements!"}
\ec
%\vskip 0.5cm
Yes, dear reader, that's what I'm telling you. Well, actually, instead of divine inspiration you might want to use a theory. Theories can sometimes constrain probability distributions in such a way that you know certain things about them {\em before} making any measurements, that is, theory can shape your {\em prior} expectation. For example, thermodynamics tells you something about the probability distribution of molecules in a container filled with gas, say, and knowing things like the temperature of that gas allows you to make predictions better than chance. But even in the absence of that, you already come equipped with {\em some} information about the system you are interested in (built into your common sense) and if you can predict with accuracy better than chance (because for example somebody told you the $p$ of the probability distribution and it is not one half), then you have some extra information. And yes, most people won't tell you that. But if you want to know about information, you first need to know.... what it is that you already know.

\section{Everything is conditional}
Let us take a few steps back for a second to contemplate the purpose of this article. 
I believe that Shannon's theory of information is a profound addition to the canon of theoretical physics. Yes, I said theoretical physics. I can't get into the details of why I think this here (but if you are wondering about this you can find my musings here~\cite{Adami2011}). But if this theory is so fundamental (as I claim) then we should make an effort to understand the basic concepts in walks of life that are not strictly theoretical physics. I tried this for molecular biology~\cite{Adami2004}, and evolutionary biology~\cite{Adami2012a}.  

But even though the theory of information is so fundamental to several areas of science, I find that it is also one of the most misunderstood theories. It seems, almost, that {\em because} ``everybody knows what information is", a significant number of people (including professional scientists) use the word, but do not bother to learn the concepts behind it.  But you really have to. You end up making terrible mistakes if you don't. The theory of information, in the end, teaches you to think about knowledge, and prediction. I'm trying to give you the entry ticket to all that. 

Here's the quick synopsis of what we have learned in the first two sections:

1.) It makes no sense to ask what the entropy of any physical system is. Because technically, it is infinite. It is only when you specify {\em what} questions you will be asking (by specifying the measurement device that you will use in order to determine the state of the random variable in question) that entropy (a.k.a. uncertainty) is finite, and defined.

2.) When you are asked to calculate the entropy of a mathematical (as opposed to physical) random variable, you are usually handed a bunch of information you didn't realize you have. Like, what's the number of possible states to expect, what those states are, and possibly even what the likelihood is of experiencing those states. But given those, your prime directive is to predict the state of the random variable as accurately as you can. And the more information you have, the better your prediction is going to be.

Now that we've got these preliminaries out of the way, it seems like high time that we get to the concept of information in earnest. I mean, how long can you dwell on the concept of entropy, really? Actually, just a bit longer as it turns out. 

I think I confused you a bit in the first two sections. One time, I write that the entropy is just $\log N$, the logarithm of the number of states the system can take on, and later I write Shannon's formula for the entropy of random variable $X$ that can take on states $x_i$ with probability $p_i$ as    
\be
H(X)=-\sum_{i=1}^N p_i\log p_i \label{enter}\;. \label{ent2}
\ee 
Actually, to be perfectly honest, I didn't even write that formula. I wrote one where there are only two states, that is, $N=2$ in Eq.~(\ref{ent2}).
And then I went on to tell you that the expression $\log N$ was ``just a special case" of Equation~(\ref{ent}). But I think I need to clear up what happened here.

In section 2, I talked about the fact that you really are given some information when a mathematician defines a random variable. Like, for example, in Eq. (\ref{ent2}) above. If you know nothing about the random variable, you don't know the $p_i$. You may not even know the range of $i$. If that's the case, we are really up the creek, with paddle {\it in absentia}. Because you wouldn't even have any idea about how much you don't know. So in the following, let's assume that you know at least how many states to expect, that is, you know $N$.
If you don't know anything else about a probability distribution, then you have to assume that each state appears with equal probability. Actually, this isn't a law or anything. I just don't know how you would assign probabilities to states if you have zero information. Nada. You just have to assume that your uncertainty is maximal in that case. And this happens to be a celebrated principle: the ``maximum entropy principle". The uncertainty (\ref{ent2}) is maximized if $p_i=1/N$ for all $i$.  And if you plug in $p_i=1/N$ in (\ref{ent2}), you get
\be
                                                 H_{\rm max}=\log N\;.   \label{ent3}
\ee
It's that simple. So let me recapitulate. If you don't know the probability distribution, the entropy is (\ref{ent3}). If you do know it, it is (\ref{ent2}). The difference between the entropies is knowledge\footnote{The statement that the most objective prior is the maximum-entropy one (implying  that it is un-informative) should be a little bit more technical than appears here, because if the probability distribution of a random variable is known to be different than the uniform distribution, then the un-informative prior is not the uniform distribution, but one that maximizes the entropy under the constraint of the likelihood of the data given the probability distribution. This is discussed in Jaynes's book~\cite{Jaynes2003},  and cases where a seemingly un-informative (uniform) prior clashes with the underlying probability distribution to make it actually informative, are discussed in~\cite{ZhuLu2004,Neumann2007}.}
The uncertainty (\ref{ent3}) does not depend on knowledge, but the entropy (\ref{ent2}) does. On a more technical note,  Eq.~(\ref{ent3}) is really just like the entropy in statistical physics when using the microcanonical ensemble, while Eq. (\ref{ent2}) is the Boltzmann-Gibbs entropy in the canonical ensemble, where the $p_i$ are given by the Boltzmann distribution.
If you have noticed that I've been using the words ``entropy" and ``uncertainty" interchangeably, I did this on purpose because they are one and the same thing here. You should use one or the other interchangeably too. But you should never say ``information" when you don't know what you can predict with the entropy at hand.

So, getting back to the narrative, one of the entropies is conditional on knowledge, while another is not. But, you think while scratching your head, wasn't there something in Shannon's work about ``conditional entropies"? Indeed, and those are the subject of this section. The section title kind of gave it away, I'm afraid. To introduce conditional entropies more formally, and then connect to (\ref{ent2}), we first have to talk about conditional probabilities. What's a conditional probability? I know, some of you groan ``I've known what a conditional probability is since I was seven!" But even you may learn something. After all, you learned something reading this article even though you're somewhat of an expert? Right? Why else would you still be reading? 
\vskip 0.5cm
\bc
                             {\em ``Infinite patience"}, you say? Moving on. 
 \ec
A conditional probability characterizes the likelihood of an event, when another event has happened at the same time. So, for example, there is a (generally small) probability that you will crash your car. The probability that you will crash your car while you are texting at the same time is considerably higher. On the other hand, the probability that you will crash your car while it is Tuesday at the same time is probably unchanged, that is, unconditional on the ``Tuesday" variable. (Unless Tuesday is your texting day, that is.)

So, the probability of events depends on what else is going on at the same time. ``Duh", you say. But while this is obvious, understanding how to quantify this dependence is key to understanding information.  In order to quantify the dependence between ``two things that happen at the same time", we just need to look at two random variables. In the case I just discussed, one variable is the likelihood that you will crash your car, and the other is the likelihood that you will be texting. The two are not always independent, you see. The problems occur when the two occur simultaneously. You know, if this was another article (like, the type where I veer off to discuss topics relevant only to theoretical physicists) I would now begin to remind you that the concept of simultaneity is totally relative, so that the concept of a conditional probability cannot even be unambiguously defined in relativistic physics (but concepts such as ``before" and ``after" can, so that helps a lot). But this is not that article, so I will just let it go.    

OK, here we go: $X$ is one random variable (think: $p_i$ is the likelihood that you crash your car while you conduct maneuver $X=x_i$, where each $x_i$ is a particular maneuver or action).  The other random variable is $Y$. That variable has only two states: either you are texting ($Y=1$), or you are not ($Y=0$) And those two states have probabilities $q_1$ (texting)  and $q_0$ (not texting)  associated with them. 
I can then write down the formula for the uncertainty of crashing your car while texting, using the probability distribution
\be
                                                    P(X=x_i|Y=1)\; .
\ee
This you can read as ``the probability that random variable $X$ is in state $x_i$ given that, at the same time, random variable $Y$ is in state $Y=1$."  This vertical bar ``$|$'' is always read as ``given".

So, let me write  $P(X=x_i|Y=1)$ as $p(i|1)$. It's much simpler that way. I can also define $P(X=x_i|Y=0)=p(i|0)$. $p(i|1)$ and $p(i|0)$ are two probability distributions that may be different (but they don't have to be if my driving is unaffected by texting). Fat chance for the latter, by the way. 

I can then write the entropy while texting as
\be
                               H(X|{\rm texting})=-\sum_{i=1}^N p(x_i|1)\log p(x_i|1)\;.  \label{cond1}
\ee
On the other hand, the entropy of the driving variable while {\em not} texting is 
\be
                           H(X|{\rm not\  texting})=-\sum_{i=1}^N p(x_i|0)\log p(x_i|0)\;.  \label{cond2}
\ee
Now, compare Eqs. (\ref{cond1}) and (\ref{cond2}) to Eq. (\ref{ent2}). The latter two are conditional entropies, conditional in this case on the co-occurrence of another event, here texting. They look just like the Shannon formula for entropy, which I told you was the one where ``you already knew something", like the probability distribution. In the case of (\ref{cond1}) and (\ref{cond2}), you know exactly what it is that you know, namely whether random variable $X$ is texting while driving, or not. 

So here's the gestalt idea that I want to get across. Probability distributions are born being uniform. In that case, you know nothing about the variable, except perhaps the number of states it can take on. Because if you didn't know {\em that}, then you wouldn't even know how much you don't know. That would be the ``unknown unknowns", that a certain political figure once injected into the national discourse. 

These probability distributions become non-uniform (that is, some states are more likely than others) once you acquire information about the states. This information is manifested by conditional probabilities. You really only know that a state is more or less likely than the random expectation if you at the same time know something else (like in the case discussed, whether the driver is texting or not). Put in another way, what I'm trying to tell you here is that any probability distribution that is not uniform (same probability for all states) is necessarily conditional. When someone hands you such a probability distribution, you may not know what it is conditional about. But I assure you that it is conditional. I'll state it as a theorem:
\vskip 0.5cm
\bc
{\bf All  probability distributions that are not uniform are in fact conditional probability distributions.}
\ec
%\vskip 0.5cm
This is not what your standard textbook will tell you, but it is the only interpretation of ``what do we know" that makes sense to me. ``Everything is conditional" thus, as the title of this section promised.

We can also write down what the {\em average} uncertainty for crashing your car is, given your texting status. It is simply the average of the uncertainty while texting and the uncertainty while not texting, weighted by the probability that you engage in any of the two behaviors.  Thus, the conditional entropy $H(X|Y)$, that is the uncertainty of crashing your car given your texting status, is
\be
                           H(X|Y)=q_0H(X|Y=0)+q_1H(X|Y=1)\;.   \label{condav}
\ee
That's obvious, right? $q_0$ being the probability that you are texting while executing any maneuver $i$, and $q_1$ the probability that you are not (while executing any maneuver).
With this definition of the entropy of one random variable given another, we can now finally tackle information.

\section{Information}
Before going on, let me quickly summarize the take-home points of sections 1-3.

1.) Entropy, also known as ``uncertainty", is something that is mathematically defined for a ``random variable". But physical objects aren't mathematical. They are messy complicated things. They become mathematical when observed through the looking glass of a measurement device that has a finite resolution. We then understand that a physical object does not ``have an entropy". Rather, its entropy is defined by the measurement device I choose to examine it with.  Information theory therefore is a theory of the relative state of measurement devices. 

2.) Entropy, also known as uncertainty, quantifies how much you don't know about something (a random variable). But in order to quantify how much you don't know, you have to know {\em something} about the thing you don't know. These are the hidden assumptions in probability theory and information theory. These are the things you didn't know you knew.

3.) Shannon's entropy is written in terms of ``$p \log p$", but these ``$p$" are really conditional probabilities if you know that they are {\em not} uniform, that is, all $p$ the same for all states. They are not uniform given what else you know. %Like, that they are not uniformly distributed. Duh, again.

I previously defined the unconditional entropy, which is the one where we know nothing about the system that the random variable describes. We call that $H_{\rm max}$, because an unconditional entropy must be maximal: it tells us how much we don't know if we don't know anything except how many states my measurement device has. Then there is the conditional entropy $H=-\sum_i p_i\log p_i$, where the $p_i$ are conditional probabilities. They are conditional on some knowledge. Thus, $H$ tells you what remains to be known. So finally, I give you:
\vskip 0.5cm
\bc
Information is: ``What you don't know minus what remains to be known given what you know". 
\ec
There it is. Clear?
\vskip 0.5cm
\bc
{\em ``Hold on, hold on. Hold on for just a minute."}
\ec
%\vskip 0.5cm
What?
\vskip 0.5cm
\bc
{\em ``This is not what I've been reading in textbooks."}
\ec
So tell me what it is that you read.
\vskip 0.5cm
\bc
{\em ``It says there that the mutual information is the difference between the entropy of random variable $X$,  $H(X)$, and the conditional entropy $H(X|Y)$, which is the conditional entropy of variable $X$ given you know the state of variable $Y$. Come to think of it, you yourself defined that conditional entropy at the end of section 3. I think it is Equation (\ref{condav}) there!" And there is this Venn diagram on Wikipedia. It looks like Figure 2!"}
\ec
%\vskip 0.5cm
\begin{figure}[htbp] %  figure placement: here, top, bottom, or page
   \centering
   \includegraphics[width=4in]{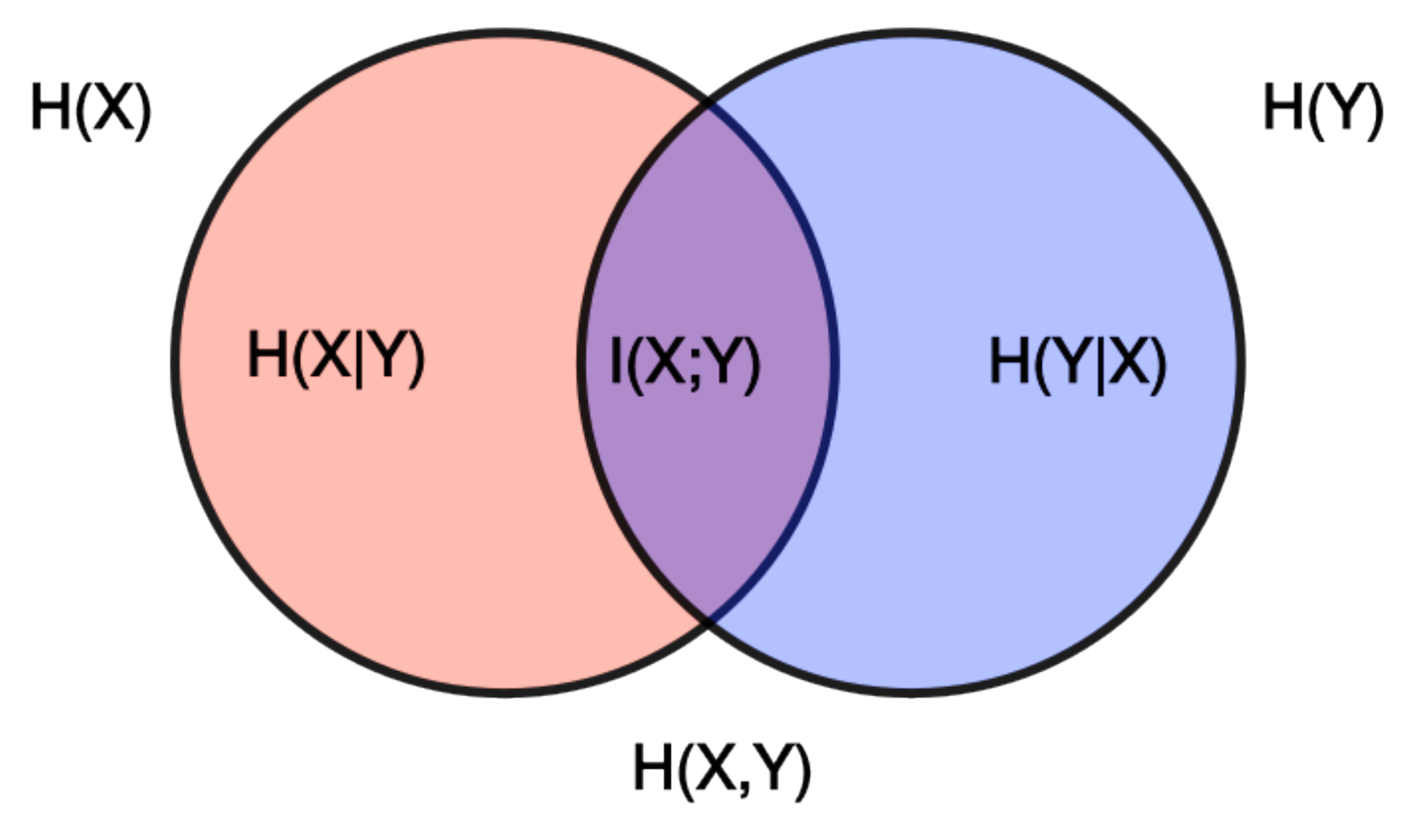} 
   \caption{Entropy Venn diagram, showing conditional and mutual entropies of two variables. Source: Wikimedia}
   \label{fig:venn}
\end{figure}

Ah, yes. That's a good diagram. Two variables $X$ and $Y$. The red circle represents the entropy of $X$, the blue circle the entropy of $Y$. The purple thing in the middle is the shared entropy $I(X:Y)$, which is what $X$ knows about $Y$. Also what $Y$ knows about $X$. They are the same thing.
\vskip 0.5cm
\bc
{\em ``You wrote $I(X:Y)$ but Wiki says $I(X;Y)$. Is your semicolon key broken?"}
\ec
Actually, there are two notations for the shared entropy (a.k.a information) in the literature. One uses the colon, the other the semicolon. Thanks for bringing this up. It confuses people. In fact, I wanted to bring up this other thing....
\vskip 0.5cm
\bc {\em ``Hold on again. You also keep on saying ``shared entropy" when Wiki says ``shared information". You really ought to pay more attention."}
\ec
%\vskip 0.5cm
Well, you. That's a bit of a pet-peeve of mine. Just look at the diagram above. The thing in the middle, the purple area, it's a shared entropy. Information is shared entropy. ``Shared information" would be, like, shared shared entropy. I think that's a bit ridiculous, don't you think?
\vskip 0.5cm
\bc
{\em ``Well, if you put it like this. I see your point. But why do I read `shared information' everywhere?"}
\ec
%\vskip 0.5cm
That is, dear reader, because people are confused about what to call entropy, and what to call information. A sizable fraction of the literature calls what we have been calling ``entropy" (or uncertainty) ``information". You can see this even in the book by Shannon and Weaver~\cite{ShannonWeaver1949} (which, come to think of it, was edited by Weaver, not Shannon). When you do this, then what is shared by the ``informations" is ``shared information". But that does not make any sense, right?
\vskip 0.5cm
\bc
{\em ``I don't understand. Why would anybody call entropy `information'? Entropy is what you don't know, information is what you know. How could you possibly confuse the two?"}
\ec
%\vskip 0.5cm
I'm with you there. Entropy is ``potential information". It quantifies ``how much you could {\em possibly} know". But it is not what you {\em actually} know. I think, between you and me, that it was just sloppy writing at first, which then ballooned into a massive misunderstanding. Both entropy and information are measured in bits, and so people would just flippantly say: ``a coin has two bits of information", when they mean to say ``two bits of entropy". And it's all downhill from there. 

I think I've made my point here, I hope. Being precise about entropy and information really matters. Colon vs. semicolon does not. Information is ``unconditional entropy minus conditional entropy". When cast as a relationship between two random variables $X$ and $Y$, we can write it as
\be
I(X:Y)=H(X)-H(X|Y)\;.
\ee
And because information is symmetric in the one who measures and the one who is being measured (remember: ``a theory of the relative state of measurement devices") this can also be written as
\be
I(X:Y)=H(Y)-H(Y|X)\;.
\ee
And both formulas can be verified by looking at the Venn diagram above.
\vskip 0.5cm
\bc
{\em ``OK, this is cool."}
\ec
%\vskip 0.5cm
\bc {\em "Hold on, hold on!"}
\ec
What is it again?
\vskip 0.5cm
\bc
{\em ``I just remembered. This was all a discussion that came after I brought up that information was $I(X:Y)=H(X)-H(X|Y)$, while you said it was $H_{\rm max}-H$, where the $H$ was clearly an entropy that you write as $H=-\sum_i p_i\log p_i$. All you have to do is look back a few pages, I'm not dreaming this!"}
\ec
So you are saying that textbooks say
\be
I=H(X)-H(X|Y) \label{info1}
\ee
while I write instead
\be
I=H_{\rm max}-H(X)\;,   \label{info2}
\ee
where $H(X)=-\sum_i p_i\log p_i$. Is that what you're objecting to?
\vskip 0.5cm
\bc
{\em 
``Yes. Yes it is."}
\ec
Well, here it is in a nutshell. In (\ref{info1}), information is defined as the difference between the actual observed entropy of $X$, minus the actual observed entropy of $X$ given that I know the state of $Y$ (whatever that state may be). 
In (\ref{info2}), information is defined as what I don't know about $X$ (without knowing any of the things that we may implicitly know already), and the actual uncertainty of $X$, given a particular probability distribution that is non-uniform. The latter entropy does not mention a system $Y$. It quantifies my knowledge of $X$ without stressing {\em what it is} that I know about $X$. If the probability distribution with which I describe $X$ is not uniform, then I do know something about $X$. My $I$ in Eq. (\ref{info2}) quantifies that. Eq. (\ref{info1}) quantifies what I know about $X$ above and beyond what I already know via Eq. (\ref{info2}), namely using my knowledge of $Y$. It quantifies specifically the information that $Y$ conveys about $X$. So you could say that the total information that I have about $X$, given that I also know the state of $Y$, would be
\be
I_{\rm total}=H_{\rm max}-H(X) + H(X)-H(X|Y)=H_{\rm max}-H(X|Y)\;.
\ee
So the difference between what I would write and what textbooks write is really only in the unconditional term: it should be maximal. But in the end, Eqs. (\ref{info1}) and (\ref{info2}) simply refer to different informations. Eq.~(\ref{info2}) is information, but I may not be aware how I got into possession of that information. Eq.~(\ref{info1}) tells me exactly the source of my information: the variable $Y$. Is it clear now?
\bc
{\em ``I'll have to get back to you on that. I'm still reading. I think I have to read it again. It sort of takes some getting used to."}
\ec
I know what you mean. It took me a while to get to that place. But, as I hinted at in the introduction, it pays off big time to have your perspective adjusted, so that you know what you are talking about when you say ``information". I have been (and will be) writing a good number of articles that reference ``information", and many of those are a consequence of research that was only possible when you understand the concept precisely. I wrote a series of articles on information in black holes already~\cite{AdamiVerSteeg2014,AdamiVerSteeg2015,BradlerAdami2014,BradlerAdami2015}. That's just the beginning. There are others, for example on how to measure how much information is stored in DNA~\cite{AdamiCerf2000} or proteins~\cite{GuptaAdami2015}, and the relationship between information and cooperation~\cite{Iliopoulosetal2010,AdamiHintze2013,HintzeAdami2015} (I mean, how you can fruitfully engage in the latter only if you have the former), and information processing in the brain~\cite{Edlundetal2011,Marstalleretal2013}. And there are more to come, on information in DNA binding sites~\cite{CliffordAdami2015} or even how to use information theory to estimate the likelihood of a spontaneous origin of life~\cite{Adami2015}. I know it sounds more like a threat than a promise. I really mean it to be a promise. 

\section{Epilogue}A kind reviewer brought to my attention an elegant and insightful article by E.T. Jaynes~\cite{Jaynes1965}, in which Jaynes not only accurately characterizes the difference between Boltzmann and Gibbs thermodynamic entropies and derives thermodynamics' second law, but also makes essentially the same statement about entropy that I have made here, by writing: 

``From this we see that entropy is an anthropomorphic concept, not only in the well-known statistical sense that it measures the extent of human ignorance as to the microstate. {\em Even at the purely phenomenological level, entropy is an anthropomorphic concept.} For it is a property, not of the physical system, but of the particular experiments you or I choose to perform on it."

While this statement pertained to thermodynamics entropy, it applies just as well to Shannon entropy as the two are intimately related. Jaynes dissects this relationship cogently in his article and I see no need to repeat it here as my focus is on information, not entropy. I have written about the relationship between thermodynamic and Shannon entropy~\cite{Adami2011} but wish I would have known Ref.~\cite{Jaynes1965} then. Perhaps this link is best summarized by noting that thermodynamics is a special case of information theory where ``all the fast things have happened, but the slow things have not" (as Richard Feynman described thermodynamical equilibrium~\cite{Feynman1972}). In other words,  in equilibrium information about the fast things has disappeared, but there may still be information about the slow things: that 's the things we don't always know we know.  

\noindent{\bf Acknowledgememnts}
I would like to thank Julyan Cartwright for suggesting to turn my blog series on information into this comment, Eugene Koonin for lively discussions about information, evolution, and genomics, as well as three referees for insightful comments.
This work was supported in part by NSF's BEACON Center for the Study of Evolution in Action, under Contract No. DBI-0939454.

%\bibliographystyle{prsoc}
%\bibliographystyle{iopart-num}

%\bibliography{PT-refs}
%%%%%%%%%% Insert bibliography here %%%%%%%%%%%%%%

\end{document}